\documentclass[aps,pra,floatfix,superscriptaddress,preprint,showpacs]{revtex4}
\usepackage{times}
\usepackage{dcolumn}                 % Aligning table columns 
\usepackage{graphicx}
\usepackage{epsf}
\usepackage{amsmath}
\usepackage{bm}
\newcolumntype{.}{D{x}{}{-1}}

\newcolumntype{w}[1]{D{.}{.}{#1}}

\begin{document}
\preprint{Version 1.0}

\title{Nuclear vector polarizability correction to hyperfine splitting}
\author{Krzysztof Pachucki}
\affiliation{
Institute of Theoretical Physics, University of Warsaw,
Ho\.{z}a 69, 00-681 Warsaw, Poland} 

\begin{abstract}
The interaction of orbital electrons with the charge and magnetic 
moment of the nucleus polarizes it, and the detailed description 
requires a careful treatment of the nuclear vector polarizability. 
We present a complete and closed form expression 
for the resulting contribution to hyperfine splitting in light atomic systems.
\end{abstract}

\maketitle
Atomic energy levels are split due to the interaction
between the nuclear and electron magnetic moments,  $\vec\mu$ and $\vec\mu_e$ respectively.
This splitting, called the hyperfine splitting (hfs), for a nonrelativistic hydrogen-like system 
in an $S$-state is given by the Fermi contact interaction 
(in units $\hbar=c=1, e^2=4\,\pi\,\alpha$)
\begin{equation}
E_{\rm F} = -\frac{2}{3}\,\langle\psi|
\vec\mu\cdot\vec\mu_e\,\delta^3(r)|\psi\rangle
=\frac{Z\,e^2}{6}\,\frac{\psi^2(0)}{M\,m}\,g\,\vec S\cdot\vec\sigma
\,,\label{01}
\end{equation}
where
\begin{equation}
\vec \mu = \frac{Z\,e}{2\,M}\,g\,\vec S\,,\label{02}
\end{equation}
$g = (1+\kappa)/S$, and $Z\,e$ and $M$ are the charge and mass of the nucleus.
For electrons the $g$-factor is close to $2$, with a small anomaly
$\kappa \approx \alpha/(2\,\pi)$, which is neglected in Eq. (\ref{01}), and
for the nucleus we assume an arbitrary $g$-factor. To compare with high accuracy measurements 
of hyperfine splitting, such as in the deuterium atom \cite{deut} ($\nu=E/h$) 
\begin{equation}
\nu_{\rm exp} = 327\,384.352\,522\,2(17)\;{\rm kHz}, \label{03}
\end{equation} 
many higher order relativistic and quantum electrodynamics (QED)
corrections have been calculated with high precision \cite{eides}.
However, the theoretical accuracy is limited by nuclear effects.
For example, in deuterium \cite{friar, sgk}
\begin{equation}
\frac{(\nu_{\rm exp}-\nu_{\rm QED})}{\nu_{\rm F}} = 138\,{\rm ppm} \label{04},
\end{equation}
where $\nu_{\rm QED}$ is the QED prediction assuming a point
nucleus. This difference is many orders of magnitude larger
than the precision of $\nu_{\rm exp}$ and $\nu_{\rm QED}$, and it
is attributable to the nuclear structure correction.
We note that the leading order $O(Z\,\alpha)$ relativistic correction 
vanishes for  a point-like and infinitely heavy nucleus. 
The subject of this work is  the detailed study of the effect of both  
finite nuclear mass and nuclear structure, which make this $O(Z\,\alpha)$ correction 
nonvanishing. This correction can be represented 
by the two-photon exchange forward scattering amplitude, which in the temporal gauge $A^0=0$
takes the form ($\omega\equiv k^0$)
\begin{equation}
\delta E_{\rm hfs} = \frac{i}{2}\,\int\frac{d\omega}{2\,\pi}\,
\int\frac{d^3k}{(2\,\pi)^3}\,\frac{1}{(\omega^2-k^2)^2}\,
\biggl(\delta^{ik}-\frac{k^i\,k^k}{\omega^2}\biggr)\,
\biggl(\delta^{jl}-\frac{k^j\,k^l}{\omega^2}\biggr)\,
t^{ji}\,T^{kl}\,\psi^2(0)\,,\label{05}
\end{equation}
where 
\begin{eqnarray}
t^{ji} &=& e^2\biggl[
\langle\bar u(p)|\gamma^j\frac{1}{\not\!p\;- \not\!k-m}\,\gamma^i|u(p)\rangle +
\langle\bar u(p)|\gamma^i\frac{1}{\not\!p\;+ \not\!k-m}\,\gamma^j|u(p)\rangle
\biggr] \nonumber \\
&=& e^2\,i\,\omega\,\epsilon^{ijk}
\,\sigma^k\,\frac{2\,(\omega^2-k^2)}
{(\omega^2-2\,m\,\omega-k^2)\,(\omega^2+2\,m\,\omega-k^2)}\,,\label{06}
\end{eqnarray}
$p$ is the momentum at rest, 
$T^{kl}$ is the corresponding virtual Compton scattering amplitude off the
nucleus, and a subtraction of the linear divergence
at small $k$ in Eq. (\ref{05}), which is related to the leading Fermi interaction,
is assumed implicitly.

One splits the nuclear structure correction $\delta E_{\rm hfs}$ into three parts:
\begin{equation}
\delta E_{\rm hfs} = \delta E_{\rm Low} + \delta E_{\rm Zemach} + \delta E_{\rm pol}.
\label{07}
\end{equation}
$\delta E_{\rm Low}$ is the leading correction to hfs of order
$Z\,\alpha\,m\,r_N$, where $r_N$ is the size of the nucleus.
This correction has been first derived by Low in \cite{low}, and has been recently 
reanalyzed and calculated for such nuclei as D, T, and $^3$He by Friar and Payne in Ref. \cite{friar}.
In this work we present a complete derivation of the Low correction,
as well as the higher order $Z\,\alpha\,m/m_p$ correction which comes from
nuclear excitations and recoil, which we denote by $\delta E_{\rm pol}$. 
The second term in Eq. (\ref{07}), $\delta E_{\rm Zemach}$ is the so called Zemach 
correction from individual nucleons. For the hydrogen atom it is
\begin{equation}
\frac{\delta E_{\rm Zemach}}{E_{\rm hfs}} = \frac{2\,\alpha\,m}{\pi^2}\,
\int\frac{d^3k}{k^4}\,\biggl[\frac{G_E(-k^2)\,G_M(-k^2)}{1+\kappa}-1\biggr]
= -2\,\alpha m\,r_{\rm Z}
\,,\label{08}
\end{equation}
where $G_E$ and $G_M$ are the electric and magnetic formfactors of the proton.
It is convenient to rewrite this correction in terms of the Zemach radius $r_{\rm Z}$
\begin{equation}
r_{\rm Z} = \int d^3 r\,\rho_E(r)\,\rho_M(r)\,r,
\end{equation}
with $\rho_E$ and $\rho_M$ being the Fourier transforms of $G_E$ and $G_M/(1+\kappa)$.
The results of Ref. \cite{friar} for the proton and neutron Zemach radius are 
$1.086(12)$ fm and $-0.042$ fm respectively.
For an arbitrary nucleus this correction is a coherent sum
of Zemach corrections from all nucleons,
\begin{equation}
\delta E_{\rm Zemach} = \frac{e^2}{6}\,\frac{\psi^2(0)}{m_p\,m}\,
\vec\sigma\cdot(-2\,\alpha\,m)\,
\bigl\langle\sum_a g_a\,\vec s_a\,r_{a\rm Z}\bigr\rangle.
\end{equation}
One notes that it would be not accurate to only use the Zemach formula with 
the nuclear elastic formfactors.
Instead one should assume a point nucleus for the QED part of hfs,
and at the first approximation
calculate $\delta E_{\rm Low}$ and $\delta E_{\rm Zemach}$ as was
done for example in Ref. \cite{friar}. 
It is still an open issue as to the accuracy of the  
elastic formfactor treatment of hfs in heavy atoms or ions,
and its relation to $\delta E_{\rm Low}$, but this problem is
not studied here.

We derive below
a complete closed form expression for $\delta E_{\rm pol}$ which can be used
to improve theoretical predictions for hyperfine splitting in light atoms or ions.
The main idea behind this derivation is the existence of an expansion
parameter in the effective nuclear Hamiltonian, 
namely the ratio of the characteristic momentum $Q$ of a nucleon to its mass $m_p$,
which is about $0.10 - 0.15$ in typical nuclei \cite{friar}.
$\delta E_{\rm pol}$ accounts for all the nuclear structure corrections,
which are $Q/m_p$ smaller than the leading $\delta E_{\rm Low}$ contribution.
To carry this out we split the integral
in Eq. (\ref{05}) into two parts. In the low energy part,
where $k$ is of order of the binding energy per nucleon, both the elastic
contribution and nuclear
excitations play a role. In the high energy part, where $k\gg$ binding energy,
nucleons are seen by the electron as free particles, and their binding energy
can simply be neglected. This derivation will be similar to the one
presented by Khriplovich {\em et al.} in Refs. \cite{khr1, khr2} for the 
particular case of the deuterium atom,
but differs in many details and in the resulting formula.
In particular we show that the leading logarithmic contribution vanishes,
while the result of Ref. \cite{khr2} can not be rewritten to such form. 

We start the derivation by noting that according to Eq. (\ref{05}),
only the antisymmetric part of the Compton amplitude $T^{ij}$ contributes to
hfs. If the scattered photon is on mass shell, $\omega = |\vec k|$, then
this amplitude $T^{ij} = i\,\omega^2\,\alpha^{ij}$ can be expressed in terms
of the vector polarizability $\alpha^{ij}$ \cite{landau}.
In the general case $\omega \neq |\vec k|$, and for convenience we will identify
the antisymmetric part of $T^{ij}$ with the vector polarizability
\begin{equation}
\alpha^{ij} = (T^{ij}-T^{ji})/2.
\end{equation}
Then using Eq. (\ref{06}), the nuclear structure correction to hfs is
\begin{eqnarray}
\delta E_{\rm hfs} &=& -\,e^2\,\psi^2(0)\,\int\frac{d\,\omega}{2\,\pi}\,
\int\frac{d^3k}{(2\,\pi)^3}\,
\frac{\bigl(\omega^2\,\epsilon^{klj}+k^i\,k^k\,\epsilon^{lij}
-k^i\,k^l\,\epsilon^{kij}\bigr)\,\sigma^j\,\alpha^{kl}}{\omega\,(\omega^2-k^2)\,
(\omega^2-2\,m\,\omega-k^2)\,(\omega^2+2\,m\,\omega-k^2)}\,,\label{08p}
\end{eqnarray}
where the Feynman integration contour is assumed and the 
apparent $1/\omega$ singularity cancels out with the numerator. 
For small photon momenta the elastic part of the vector polarizability
\begin{equation}
\alpha^{ij} = 
i\,\epsilon^{ijk}\,\frac{(Z\,e)^2}{M^2\,\omega}\,
\Bigl[\omega^2\,S^k\,(g-1)
-\vec k^2\,S^k\,g/2
-k^k\,(\vec k\cdot\vec S)\,g\,(g-2)/4\Bigr]\,,\label{09}
\end{equation}
can be obtained from the following Hamiltonian,
which describes the interaction of a particle having  charge $Z\,e$,  mass $M$, 
and spin $S$ with the electromagnetic field
\begin{equation}
H = \frac{\vec\Pi^2}{2\,M} + Z\,e\,A^0 -\frac{Z\,e}{2\,M}\,g\,\vec S\cdot\vec B
-\frac{Z\,e}{4\,M^2}\,(g-1)\,\vec S\cdot[\vec E\times\vec\Pi-\vec\Pi\times\vec E].\label{10}
\end{equation}
where $\vec\Pi$ is defined in Eq. (\ref{14}). 
Eq. (\ref{09}) is in fact a low energy virtual Compton scattering amplitude
off a point nucleus with arbitrary spin $S$. It agrees with results obtained 
for the first time by Khriplovich {\em et al} in Ref. \cite{khr1}. 
To obtain the corresponding contribution to the hyperfine splitting 
$\delta_0 E_{\rm hfs}$, we assume
$|\vec k|<\Lambda$, with the cut-off $\Lambda$ being larger than
binding energy per nucleon, and smaller than the inverse of the nuclear size.
We will show later that the dependence on $\Lambda$ cancels out
between all the corrections, so there is no need to specify its precise value.
After  subtraction of the leading nonrelativistic part $\omega\sim k^2$,
which is a Coulomb iteration of the Fermi contact interaction, and after the
$\omega$ and $k$ integrations of $\delta_0E_{\rm hfs}$ are performed, it takes the form
\begin{equation}
\delta_0E_{\rm hfs} = 
e^2\,(Z\,e)^2\,\psi^2(0)\,\frac{\vec\sigma\cdot\vec S}{M^2}\,
\frac{1}{64\,\pi^2}\,\biggl[
\ln\biggl(\frac{2\,\Lambda}{m}\biggr)\,(g^2-4\,g-12)+
\frac{1}{6}\,(g^2+124\,g+4)\biggr].\label{11}
\end{equation}
The important point of this calculation is that after 
expansion in the small parameter $m/\Lambda$ or $\Delta E/\Lambda$,
the limit $\Lambda\rightarrow 0$ is performed consistently
in all the parts. 

Apart from the elastic contribution, various nuclear excitations
play a role, even for small values of the photon momentum $k$.
Their calculation is more complicated, as the nuclear 
Hamiltonian is not well understood. We will use 
the interaction Hamiltonian with the electromagnetic field
which was obtained by us in \cite{krp1}. The main assumption is 
that the characteristic wavelength of the electromagnetic field is much larger 
than the nuclear size. Although it was derived for a system
consisting of electromagnetically interacting particles,
we assume that the obtained form should be valid also for 
nucleons: this follows from the gauge and Lorentz symmetries
of the nucleon-nucleon interactions. However, this assumption
can be explicitly verified only by using a systematic method to 
derive the nuclear Hamiltonian such as chiral perturbation theory ($\chi$PT), but this will not be done here. 
This Hamiltonian, dropping  terms that do not contribute 
to the vector polarizability, is \cite{krp1}
\begin{eqnarray}
H &=& H_{\rm IN} + \frac{\vec\Pi^{\,2}}{2\,M} + Z\,e\,A^0 -
\frac{Z\,e}{2\,M}\,g\,\vec S\cdot\vec B +  
\frac{Z\,e}{4\,M^2}\,(g-1)\,\vec S\cdot
\bigl(\vec\Pi\times\vec E-\vec E\times\vec\Pi\bigr)\nonumber \\ &&
-\biggl[\sum_a\frac{e_a}{2\,m_a}\,
(\vec l_a+g_a\,\vec s_a)-\frac{Z\,e}{2\,M}\,g\,\vec S\biggr]\cdot\vec B
-\sum_a e_a\,\vec x_a\biggl(\vec E+\frac{1}{2\,M}\,\vec\Pi\times\vec B
-\frac{1}{2\,M}\,\vec B\times\vec\Pi\biggr)\nonumber \\ &&
+\sum_a\biggl\{
-\frac{e_a}{2}\,(x^i_a\,x^j_a-x_a^2\,\delta^{ij}/3)\,E^i_{,j}
-\biggl[\frac{e_a}{2\,m_a}(g_a-1)-\frac{Z\,e}{2\,M}\biggr]\vec
s_a\times\vec x_a\cdot\partial_t\vec E
 \nonumber \\ && 
-\frac{e_a}{6\,m_a}\,(l_a^j\,x_a^i+x_a^i\,l_a^j)\,B^j_{,i}
-\frac{e_a}{2\,m_a}\,g_a\,x_a^i\,s_a^j\,B^j_{,i}
+\frac{Z\,e}{6\,M}\,(\vec l_a\times\vec x_a -\vec x_a\times\vec l_a)\cdot\partial_t \vec E\biggr\}.
\label{12}
\end{eqnarray}
where the index $a$ goes over protons and neutrons, 
$M=\sum_a m_a$ and $Z\,e = \sum_a e_a$.
For protons $e_a = e$, $g_a = g_p$, while for neutrons $e_a\,g_a = e\,g_n$,
but alone $e_a\rightarrow 0$. The electromagnetic field and its derivatives
are assumed in the above equation, to be at point $\vec R$. Center of mass coordinates are defined by
\begin{eqnarray}
\vec R &=& \frac{1}{M}\,\sum_a m_a\,\vec r_a\,,\label{13}\\
\vec \Pi &=& \sum_a \vec p_a-e_a\,\vec A(R) = \vec P-Z\,e\,\vec A(R)\,.\label{14}
\end{eqnarray}
The relative coordinates, which are defined by
\begin{eqnarray}
\vec x_a &=& \vec r_a-\vec R\,,\label{15}\\
\vec q_a &=& \vec p_a-\frac{m_a}{M}\,\vec P\,,\label{16}
\end{eqnarray}
obey nonstandard commutation relations
\begin{equation}
[x_a^i\,,\,q_b^j] = i\,\delta^{ij}\,\biggl(\delta_{ab}-\frac{m_b}{M}\biggr).\label{17}
\end{equation}
The total spin $\vec S$ is
\begin{equation}
\vec S = \sum_a \vec s_a + \vec l_a,\label{18}
\end{equation}
where the orbital angular momentum $l_a$ is
\begin{equation}
\vec l_a = \vec x_a\times\vec q_a\,,\label{19}
\end{equation}
and the $g$-factor is defined by
\begin{equation}
\frac{Z\,e\,g}{2\,M}\,\vec S \equiv \Bigl\langle\sum_a 
\frac{e_a}{2\,m_a}\,(\vec l_a+g_a\,\vec s_a)\Bigr\rangle\,.\label{20}
\end{equation}
$H_{\rm IN}$ is the internal  Hamiltonian of the nucleus, the exact form of which is not well known.
For further calculations we will assume that the following commutator 
\begin{equation}
[H_{\rm IN}\,,\,x_a^i] = -i\,\frac{q_a^i}{m_a},\label{21}
\end{equation}
holds, at least to a good approximation. 

Since the nuclear excitation
energies are larger than the electron mass $m$ (with a few exceptions), 
we neglect $m$ in Eq. (\ref{08p}) and obtain
\begin{eqnarray}
\delta E_{\rm hfs} &=& -e^2\,\psi^2(0)\,\int\frac{d\,\omega}{2\,\pi}\,
\int^\Lambda\frac{d^3k}{(2\,\pi)^3}\,
\frac{\bigl(\omega^2\,\epsilon^{klj}+k^i\,k^k\,\epsilon^{lij}
-k^i\,k^l\,\epsilon^{kij}\bigr)
\,\sigma^j\,\alpha^{kl}}{\omega\,(\omega^2-k^2)^3}.\label{22}
\end{eqnarray} 
There are various contributions to $\alpha^{kl}$ which follow from
the Hamiltonian in Eq. (\ref{12}),  and we calculate them in order.
The first contribution is due to the electric dipole coupling
\begin{equation}
\delta H = -\sum_a e_a\,\vec x_a\cdot\vec E \equiv -\vec D\cdot\vec E\,.\label{23}
\end{equation}
This contribution has already been considered by Friar and Payne in Ref. \cite{friar2}
for the particular case of the deuterium atom. For the general nucleus
the effect on the vector polarizability is analogous
\begin{equation}
\delta \alpha^{kl} = \omega^2\,
\biggl\langle D^k\frac{1}{E-H_{\rm IN}-\omega}\,D^l +
D^l\frac{1}{E-H_{\rm IN}+\omega}\,D^k\biggr\rangle\,,\label{24}
\end{equation} 
and the corresponding contribution to the hyperfine splitting is
\begin{equation}
\delta_1 E_{\rm hfs} =
-\frac{3\,i}{16\,\pi^2}\,\,e^2\,\psi^2(0)\,\epsilon^{ijk}\,\sigma^k\,
\bigl\langle D^i\,\ln[2\,(H_{\rm IN}-E)/m]\,D^j \bigr\rangle\,.\label{25}
\end{equation}
The constant $m$ (the electron mass) in the argument 
of the logarithm is not relevant here, 
as it does not affect the above matrix element, since $D^i$ commutes with $D^j$.
In the nonrelativistic approximation $\delta_1 E_{\rm hfs}$ 
vanishes. Only the presence of relatively small 
spin-orbit terms in $H_{\rm IN}$ causes it  to be nonvanishing,
therefore this correction $\delta_1 E_{\rm hfs}$ is not expected to be 
the most significant one.

The second contribution is due to magnetic dipole coupling in Eq. (\ref{12})
\begin{equation}
\delta H =-\biggl[\sum_a\frac{e_a}{2\,m_a}\,
(\vec l_a+g_a\,\vec s_a)-\frac{Z\,e}{2\,M}\,g\,\vec S\biggr]\cdot\vec B 
\equiv -(\vec \mu-\langle\vec \mu\rangle)\cdot\vec B.\label{26}
\end{equation} 
The corresponding vector polarizability is
\begin{eqnarray}
\delta \alpha^{kl} &=&
\biggl\langle\bigl[(\vec\mu-\langle\vec\mu \rangle)\times\vec k\bigr]^k
\frac{1}{E-H_{\rm IN}-\omega}\,
\bigl[(\vec\mu-\langle\vec\mu \rangle)\times\vec k\bigr]^l
\nonumber \\ && +
\bigl[(\vec\mu-\langle\vec\mu \rangle)\times\vec k\bigr]^l\frac{1}{E-H_{\rm IN}+\omega}
\,\bigl[(\vec\mu-\langle\vec\mu \rangle)\times\vec k\bigr]^k\biggr\rangle,\label{27}
\end{eqnarray}
and the contribution to the hyperfine splitting is
\begin{eqnarray}
\delta_2 E_{\rm hfs} &=&
\frac{1}{16\,\pi^2}\,\,e^2\,\psi^2(0)\,\sigma^k\,\biggl\{i\,\epsilon^{ijk}\,
\langle(\mu^i-\langle\mu^i\rangle)\,\ln[2\,(H_{\rm IN}-E)/m]\,(\mu^j-\langle\mu^j\rangle)\rangle
\label{28} \\ &&
+\biggl[\ln\biggl(\frac{m}{2\,\Lambda}\biggr)+\frac{4}{3}\biggr]\,
\biggl\langle\biggl(\frac{e\,g}{2\,M}\biggr)^2\,S^k
-\sum_a\biggl(\frac{e_a}{2\,m_a}\biggr)^2\bigl(g_a^2\,s_a^k+l_a^k\bigr)
+\frac{(\vec D\times\vec Q)^k}{4\,M^2}\biggr\rangle\biggr\}\,,\nonumber
\end{eqnarray}
where
\begin{equation}
\frac{\vec Q}{M} \equiv i\,[H,\vec D] = \sum_a \frac{e_a}{m_a}\,\vec q_a\,.\label{29}
\end{equation} 
 
The third contribution is due to the electric quadrupole coupling
\begin{equation}
\delta H = -\frac{1}{2}\sum_a
e_a\,(x_a^i\,x_a^j-x_a^2\,\delta^{ij}/3)\,E^i_{,j}
\equiv  -\frac{1}{2}\,D^{ij}\,E^i_{,j}\,.\label{30}
\end{equation}
The corresponding vector polarizability is
\begin{equation}
\delta \alpha^{kl} = \frac{\omega^2\,k^i\,k^j}{4}\,
\biggl\langle D^{ki}\,\frac{1}{E-H_{\rm IN}-\omega}\,D^{lj}+
D^{lj}\,\frac{1}{E-H_{\rm IN}+\omega}\,D^{ki}\biggr\rangle\,,
\label{31}
\end{equation}
and the contribution to the hyperfine splitting is
\begin{eqnarray}
\delta_3 E_{\rm hfs} &=&
-\frac{3\,i}{64\,\pi^2}\,\frac{e^2}{M^2}\,\psi^2(0)\,\sigma^k\,\epsilon^{ijk}\,
\langle Q^{i\,l}\,\ln[2\,(H_{\rm IN}-E)/m]\,Q^{j\,l}\rangle
\nonumber \\ && 
+\frac{15}{64\,\pi^2}\,e^2\,\psi^2(0)\,
\biggl[\ln\biggl(\frac{m}{2\,\Lambda}\biggr)+\frac{4}{15}\biggr]\,\vec \sigma\cdot
\biggl\langle\frac{\vec D\times\vec Q}{M^2}
-\sum_a\biggl(\frac{e_a}{m_a}\biggr)^2\,\vec l_a\biggr\rangle\biggr\},
\label{32}
\end{eqnarray}
where
\begin{equation}
\frac{Q^{ij}}{M} \equiv i\,[H, D^{ij}] = 
\sum_a \frac{e_a}{m_a}\,\bigl[x_a^i\,q_a^j+q_a^i\,x_a^j
-(\vec x_a\,\vec q_a + \vec q_a\,\vec x_a)\,\delta^{ij}/3\bigr].\label{33}
\end{equation} 

The contribution coming from the magnetic dipole on one side and 
the electric quadrupole on the other side is being neglected.
This is because the resulting matrix element
\begin{equation}
\langle Q^{kj}\,\ln[(H_{\rm IN}-E)/\Lambda]\,\mu^k-{\rm h.c}\,\rangle,\label{34}
\end{equation}
involves operators $Q^{kj}$ and $\mu^k$ which commute when 
the implicit sum over $k$ is assumed.
Therefore this matrix element does not depend on $\Lambda$, is small,
and will consequently be neglected. This argument will be used several times
in neglecting or simplifying expressions in the following. 

The next contribution comes from
relativistic corrections to the electric dipole coupling,
\begin{equation}
\delta H = -\sum_a \biggl[\frac{e_a}{2\,m_a}(g_a-1)-\frac{Z\,e}{2\,M}\biggr]\vec
s_a\times\vec x_a\cdot\partial_t\vec E\,.\label{35}
\end{equation}
The vector polarizability correction is
\begin{eqnarray}
\delta \alpha^{kl} &=& i\,\omega^3\,\sum_a
\biggl[\frac{e_a}{2\,m_a}(g_a-1)-\frac{Z\,e}{2\,M}\biggr]\,
\biggl\langle D^k\,\frac{1}{E-H_{\rm IN}-\omega}\,(\vec s_a\times\vec x_a)^l
\nonumber \\ &&
+(\vec s_a\times\vec x_a)^l\,\frac{1}{E-H_{\rm IN}+\omega}\,D^k-
(k\leftrightarrow l, \omega\rightarrow -\omega)\biggr\rangle,\label{36}
\end{eqnarray}
and the contribution to the hyperfine splitting is
\begin{eqnarray}
\delta_4 E_{\rm hfs} &=&
-\frac{i\,e^2}{16\,\pi^2\,M}\,\psi^2(0)\,\sum_a
\biggl[\frac{e_a}{m_a}(g_a-1)-\frac{Z\,e}{M}\biggr]\,\sigma^i\,
\biggl\langle Q^k\,\ln\biggl[\frac{2\,(H_{\rm IN}-E)}{m}\biggr]\,x_a^k\,s_a^i 
- {\rm h.c.} \biggr\rangle
\nonumber \\ &&
-\frac{3\,e^2}{16\,\pi^2}\,\psi^2(0)\,
\biggl[\ln\biggl(\frac{m}{2\,\Lambda}\biggr)+\frac{8}{9}\biggr]\,\vec \sigma\cdot
\sum_a \vec s_a\biggl(\frac{e_a}{m_a}-\frac{Z\,e}{M}\biggr)
\biggl[\frac{e_a}{m_a}\,(g_a-1)-\frac{Z\,e}{M}\biggr].\label{37}
\end{eqnarray}
We have used in the above an approximate relation
\begin{equation}
\langle Q^k\,\ln [2\,(H_{\rm IN}-E)/m]\,x_a^i\,s_a^k-{\rm h.c.}\,\rangle\approx
\langle Q^k\,\ln [2\,(H_{\rm IN}-E)/m]\,x_a^k\,s_a^i-{\rm h.c.}\,\rangle/3,\label{38}
\end{equation}
which comes from the fact that the commutator
$[Q^k,x_a^i\,s_a^k - x_a^k\,s_a^i/3]$ vanishes.

In the fifth contribution one vertex remains $-\vec D\,\vec E$
but the other one is
\begin{equation}
\delta H = -\sum_a\frac{e_a}{2\,m_a}\,g_a\,x_a^i\,s_a^j\,B^j_{,i}\,.\label{39}
\end{equation}
The vector polarizability is
\begin{eqnarray}
\delta \alpha^{kl} &=& i\,\omega\,\sum_a\,\frac{e_a\,g_a}{2\,m_a}\,
\biggl\langle
\vec x_a\vec k\,(\vec s_a\times\vec k)^k\,\frac{1}{E-H_{\rm IN}-\omega}\,D^l
\nonumber \\ &&
+D^l\,\frac{1}{E-H_{\rm IN}+\omega}\,\vec x_a\vec k\,(\vec s_a\times\vec k)^k
-(k\leftrightarrow l, \omega\rightarrow -\omega)\biggr\rangle,\label{40}
\end{eqnarray}
and the corresponding correction to the hyperfine splitting is
\begin{eqnarray}
\delta_5 E_{\rm hfs} &=& -\frac{i\,e^2}{6\,\pi^2}\,\psi^2(0)\,\frac{\sigma^k}{M}\,
\sum_a\,\frac{e_a\,g_a}{2\,m_a}\,\langle s_a^k\,x_a^j\,\ln[2\,(H_{\rm IN}-E)/m]\,Q^j
-{\rm h.c.} \rangle
\nonumber \\ &&
+\frac{e^2}{2\,\pi^2}\,\psi^2(0)\,
\biggl[\ln\biggl(\frac{m}{2\,\Lambda}\biggr)+\frac{1}{3}\biggr]\,\vec\sigma\,
\sum_a\,\frac{e_a\,g_a}{2\,m_a}\,\vec s_a\,\biggl(\frac{e_a}{m_a}-\frac{Z\,e}{M}\biggr).\label{41}
\end{eqnarray}

The sixth contribution is due to the following modification
of one of the vertices
\begin{equation}
\delta H = -\sum_a \frac{e_a}{6\,m_a}\,(l_a^j\,x_a^i+x_a^i\,l_a^j)\,B^j_{,i}\,.\label{42}
\end{equation}
It is very similar to the previous one, and can be obtained by replacing
spin with the orbital angular momentum. The vector polarizability is
\begin{eqnarray}
\delta \alpha^{kl} &=& i\,\omega\,\sum_a\,\frac{e_a}{6\,m_a}\,
\biggl\langle
\bigl\{\vec x_a\vec k\,,\,(\vec l_a\times\vec k)^k\bigr\}\,\frac{1}{E-H_{\rm IN}-\omega}\,D^l
\nonumber \\ &&
+D^l\,\frac{1}{E-H_{\rm IN}+\omega}\,\bigl\{\vec x_a\vec k\,,\,(\vec l_a\times\vec k)^k\bigr\}
-(k\leftrightarrow l, \omega\rightarrow -\omega)\biggr\rangle,\label{43}
\end{eqnarray}
and the corresponding correction to the hyperfine splitting is
\begin{eqnarray}
\delta_6 E_{\rm hfs} &=& -\frac{i\,e^2}{32\,\pi^2}\,\psi^2(0)\,\frac{\sigma^k}{M}\,
\sum_a\,\frac{e_a}{m_a}\,\langle(l_a^k\,x_a^j+x_a^j\,l_a^k)\,\ln[2\,(H_{\rm IN}-E)/m]\,Q^j
-{\rm h.c.}\rangle
\nonumber \\ &&
+\frac{e^2}{4\,\pi^2}\,\psi^2(0)\,
\biggl[\ln\biggl(\frac{m}{2\,\Lambda}\biggr)+\frac{1}{3}\biggr]\,\vec \sigma \cdot
\sum_a\,\vec l_a\,\frac{e_a}{m_a}\,\biggl(\frac{e_a}{m_a}-\frac{Z\,e}{M}\biggr).\label{44}
\end{eqnarray}

The seventh contribution is due to a relativistic,  spin independent
correction to the electric dipole operator,
\begin{equation}
\delta H = \sum_a
\frac{Z\,e}{6\,M}\,(\vec l_a\times\vec x_a -\vec x_a\times\vec l_a)\cdot\partial_t \vec E\,.\label{45}
\end{equation} 
The vector polarizability can simply be obtained from Eq. (\ref{36}) 
\begin{eqnarray}
\delta \alpha^{kl} &=& -i\,\omega^3\,\frac{Z\,e}{6\,M}\,\sum_a
\biggl\langle D^k\,\frac{1}{E-H_{\rm IN}-\omega}\,(\vec l_a\times\vec x_a -\vec x_a\times\vec l_a)^l
\nonumber \\ &&
+(\vec l_a\times\vec x_a -\vec x_a\times\vec l_a)^l\,\frac{1}{E-H_{\rm IN}+\omega}\,D^k
-(k\leftrightarrow l, \omega\rightarrow -\omega)\biggr\rangle,
\label{46}
\end{eqnarray}
and the contribution to the hyperfine splitting is
\begin{eqnarray}
\delta_7 E_{\rm hfs} &=&
-\frac{3\,i}{128\,\pi^2}\,\,e^2\,\psi^2(0)\,\sigma^i\,\frac{Z\,e}{M}\,
\bigl\langle\sum_a\,(x_a^j\,l_a^i+l_a^i\,x_a^j)
\,\ln[2\,(H_{\rm IN}-E)/m]\,Q^j - {\rm h.c.}\bigr\rangle
\nonumber \\ &&
+\frac{3}{16\,\pi^2}\,e^2\,\psi^2(0)\,
\biggl[\ln\biggl(\frac{m}{2\,\Lambda}\biggr)+\frac{8}{9}\biggr]\,\,\sigma^k\,
\sum_a l_a^{\,k}\,\biggl(\frac{e_a}{m_a}-\frac{Z\,e}{M}\biggr)\,\frac{Z\,e}{M}.\label{47}
\end{eqnarray}
We have used in the above the approximate relation
\begin{eqnarray}
&& \bigl\langle\sum_a\,(x_a^i\,l_a^j+l_a^j\,x_a^i)
\,\ln[2\,(H_{\rm IN}-E)/m]\,Q^j - {\rm h.c.}\bigr\rangle\nonumber \\
&\approx&
\bigl\langle\sum_a\,(x_a^j\,l_a^i+l_a^i\,x_a^j)
\,\ln[2\,(H_{\rm IN}-E)/m]\,Q^j - {\rm h.c.}\bigr\rangle/4,\label{48}
\end{eqnarray}
which comes from the fact that the commutator 
$[Q^j,(x_a^i\,l_a^j+l_a^j\,x_a^i) - (x_a^j\,l_a^i+l_a^i\,x_a^j)/4] $ vanishes.

The remaining two low energy contributions involve the momentum of the nucleus
$\vec P$. The following correction
\begin{equation}
\delta H = -\frac{\vec D}{2\,M}\,\bigl(\vec P\times\vec B-\vec B\times\vec P\bigr),\label{53}
\end{equation}
replaces the electric dipole coupling $-\vec D\,\vec E$. The resulting
vector polarizability is 
\begin{eqnarray}
\delta \alpha^{kl} &=& \frac{\omega}{2\,M}\,(\delta^{jl}\,k^2-k^j\,k^l)\,
\biggl\langle D^k\frac{1}{E-H_{\rm IN}-\omega}\,D^j - 
D^j\frac{1}{E-H_{\rm IN}+\omega}\,D^k  \biggr\rangle
\nonumber \\ &&-(k\leftrightarrow l, \omega\rightarrow -\omega) ,\label{54}
\end{eqnarray}
and the contribution to  hyperfine splitting is
\begin{equation}
\delta_8  E_{\rm hfs} =
\frac{3\,e^2}{16\,\pi^2}\,\psi^2(0)\frac{\vec\sigma}{M^2}\, 
\bigl\langle\vec D\times\ln[2\,(H_{\rm IN}-E)/m]\,\vec Q
+\ln[m/(2\,\Lambda)]\,\vec D\times\,\vec Q\bigr\rangle.
\label{55}
\end{equation}
The kinetic energy of the nucleus
\begin{equation}
\delta H = \frac{\vec P^2}{2\,M},\label{56}
\end{equation}
modifies the vector polarizability in Eq. (\ref{24}) by adding $k^2/(2\,M)$ to $H_{\rm IN}$
which results in the following correction to the hyperfine splitting:
\begin{eqnarray}
\delta_9 E_{\rm hfs} &=&
-\frac{7\,e^2}{16\,\pi^2}\,\psi^2(0)\,\frac{\vec \sigma}{M^2}\,
\biggl\langle\vec D\times\ln[2\,(H_{\rm IN}-E)/m]\,\vec Q
+\biggl[\ln\biggl(\frac{m}{2\,\Lambda}\biggr)+\frac{29}{42}\biggr]\,
\vec D\times\,\vec Q \biggr\rangle.\label{57}
\end{eqnarray}

The last contribution comes from photon momenta $|\vec k|>\Lambda$.
Since $\Lambda$ is much larger than the binding energy per nucleon
one can  completely neglect the nucleon-nucleon interaction.
Therefore, the electron effectively sees free protons and neutrons.
In the case, when a photon is emitted and absorbed by the same nucleon, 
we replace 
\begin{equation}
\int_\Lambda dk \,(\ldots) = \int dk\,(\ldots) - \int^\Lambda dk\,(\ldots)
\label{57p}
\end{equation}
and separately consider both terms. The second term gives a contribution
similar to $\delta_{0} E_{\rm hfs}$, but with the opposite sign:
\begin{equation}
\delta_{10} E_{\rm hfs} = -\sum_a e^2\,\psi^2(0)\,\frac{\vec\sigma\cdot\vec s_a}{m_a^2}\,
\frac{e_a^2}{64\,\pi^2}\,\biggl[
\ln\biggl(\frac{2\,\Lambda}{m_a}\biggr)\,(g_a^2-4\,g_a-12)+
\frac{1}{6}\,(g_a^2+124\,g_a+4)\biggr],\label{58}
\end{equation}
while the first term of Eq. (\ref{57p}) consists of a Zemach correction 
which is already included in Eq. (\ref{07}) as $\delta E_{\rm Zemach}$, 
and a small polarizability with recoil corrections from individual nucleons,  
which we account for in  $\delta g$
\begin{equation}
\delta_{11} E_{\rm hfs} = \sum_a \frac{e^2}{6\,m\,m_a}\,\psi^2(0)\,
\vec\sigma\,\vec s_a\,\delta g_a\,.\label{59}
\end{equation}
For the proton the sum of the recoil and 
polarizability corrections, $5.84$ ppm and $1.30$ ppm respectively, amounts to \cite{carl}
\begin{equation}
\frac{\delta g_p}{g_p} = 7.14\;{\rm ppm}\,,\label{60}
\end{equation}
while for the neutron it has not yet been obtained,
although we expect it to be much smaller.

We now consider the case where the photon is emitted and absorbed by different nucleons,
denoting the resulting correction as $\delta E_{\rm Low}$.
The Compton amplitude $T^{\mu\nu}$ obeys $k_\mu\,T^{\mu\nu}=0$.
From this one obtains $k^i\,T^{ik} = \omega T^{0k}$, and uses this
to rewrite the formulae (\ref{22}) in the form
\begin{eqnarray}
\delta E_{\rm hfs} &=& -e^2\,\psi^2(0)\,\int\frac{d\,\omega}{2\,\pi}\,
\int_\Lambda\frac{d^3k}{(2\,\pi)^3}\,
\frac{1}{(\omega^2-k^2)^3}\,\sigma^j\,\epsilon^{klj}\,
\bigl[\omega\,T^{kl}+k^l\,(T^{0k}-T^{k0})\bigr].
\label{61}
\end{eqnarray} 
Since the dominating contribution in the above integral comes from
$\omega, k$ of order of the inverse of the nuclear size, which is much smaller
than the nucleon mass, we may use the nonrelativistic
approximation for $T^{\mu\nu}$,
\begin{eqnarray}
T^{\mu\nu} &=&\sum_{a\neq b}\int d^3 r\,d^3 r'\biggl\langle
J^\mu_a(r)\,e^{i\,\vec k\,\vec r}\frac{1}{E-H_{\rm IN}-\omega+i\,\epsilon}
J^\nu_b(r')\,e^{-i\,\vec k\,\vec r'} \nonumber \\ && + 
J^\nu_b(r')\,e^{-i\,\vec k\,\vec r'}\frac{1}{E-H_{\rm IN}+\omega+i\,\epsilon}
J^\mu_a(r)\,e^{i\,\vec k\,\vec r}\biggr\rangle.
\label{62}
\end{eqnarray}
The contribution coming from $T^{kl}$ in Eq. (\ref{61})
is of nominal order $Z\,\alpha\,m/m_p$, therefore 
after carrying out the $\omega$ integration one can neglect
$H_{\rm IN}-E$ in comparison to $k$. It is then
proportional to the commutator $[J^k_a(r),J^l_b(r)]$,
which vanishes for different nucleons $a\neq b$. The second term
in Eq. (\ref{61}) involving $T^{0j}$ is of nominal order
$Z\,\alpha\,m\,r_N$, therefore one shall keep also the second term
in the expansion:
\begin{equation}
\frac{1}{k+H_{\rm IN}-E} \approx \frac{1}{k} -\frac{H_{\rm IN}-E}{k^2}.
\label{63}
\end{equation}
However, this second term leads to the correction proportional to
\begin{eqnarray}
&&\bigl\langle J^0_a(r)\,e^{i\,\vec k\,\vec r}\,(H_{\rm IN}-E)\,J^i_b(r')\,e^{-i\,\vec k\,\vec r'}+
J^i_b(r')\,e^{-i\,\vec k\,\vec r'}\,(H_{\rm IN}-E)\,J^0_a(r)\,e^{i\,\vec k\,\vec r}\bigr\rangle
\nonumber \\
&=& \bigl\langle [J^0_a(r)\,e^{i\,\vec k\,\vec r},[H_{\rm IN}-E, J^i_b(r')\,e^{-i\,\vec k\,\vec r'}]]\bigr\rangle,
\label{64}
\end{eqnarray}
which also vanishes for $a\neq b$. Therefore $\delta E_{\rm Low}$ becomes
\begin{equation}
\delta E_{\rm Low} = e^2\,\psi^2(0)\,\int_\Lambda \frac{d^3 k}{(2\,\pi)^3}\,
\frac{2\,i}{k^6}\,\vec \sigma\times\vec k\,\sum_{a\neq b}\int d^3 r\,d^3 r'
\bigl\langle J_a^0(r)\,e^{i\,\vec k\,(\vec r-\vec r')}\,\vec J_b(r')\bigr\rangle.
\label{65}
\end{equation}
The linear divergence at small $\Lambda$ is eliminated by subtraction of 
$1/k^4$ term from the integrand, which is related to the leading Fermi
interaction. We should subtract it at the beginning, but for convenience
we do it now. Assuming the nonrelativistic approximation for the
interaction of the nucleon with the electromagnetic field,
\begin{equation}
H = \frac{(\vec p-e\,\vec A)^2}{2\,m} + e\,A^0 - \vec\mu\cdot\vec B, \label{66}
\end{equation}
one obtains the ``Low'' nuclear structure correction
to the hyperfine splitting \cite{low}
\begin{eqnarray}
\delta E_{\rm Low} &=& e^2\,\psi^2(0)\,\int\frac{d^3k}{(2\,\pi)^3}\,
\frac{2\,i}{k^6}\,\vec\sigma\times\vec k\,\sum_{a\neq b} \frac{e_a\,e_b}{m_b}\,
\biggl\langle e^{i\,\vec k\,\vec r_{ab}}\,
\biggl(\vec p_b-\frac{i\,g_b}{4}\,\vec\sigma_b\times\vec k\biggr)\biggr\rangle
\nonumber \\ &=&
\frac{\alpha}{16}\,\psi^2(0)\,\vec\sigma\,\sum_{a\neq b}
\frac{e_a\,e_b}{m_b}\,\biggl\langle
4\,r_{ab}\,\vec r_{ab}\times\vec p_b +\frac{g_b}{r_{ab}}
\bigl[\vec r_{ab}\,(\vec r_{ab}\cdot\vec\sigma_b)-3\,\vec\sigma_b\,r_{ab}^2 \bigr]\biggr\rangle
\label{67}
\end{eqnarray}
A more accurate treatment of $\delta E_{\rm Low}$
that includes the finite size of the nucleons 
is presented in Ref. \cite{friar}, where the authors emphasize the importance
of additional meson exchange currents.

We now present
the complete nuclear polarizability correction to the hyperfine splitting
$\delta E_{\rm pol}$, and point out that it does not depend on the artificial parameter $\Lambda$
\begin{eqnarray}
\delta E_{\rm pol} &=& \sum_{i=0}^{11} \delta_i E_{\rm hfs} =
\frac{e^2}{64\,\pi^2}\,\psi^2(0)\,\sigma^k\,\biggl\{
-12\,i\,\epsilon^{ijk}\,
\bigl\langle D^i\,{\rm Ln}\,D^j \bigr\rangle
\nonumber \\ &&
+4\,i\,\epsilon^{ijk}\,
\bigl\langle (\mu^i-\langle\mu^i\rangle)\,{\rm Ln}\,(\mu^j-\langle\mu^j\rangle)\bigr\rangle
-\frac{3\,i}{M^2}\,\epsilon^{ijk}\,
\bigl\langle Q^{i\,l}\,{\rm Ln}\,Q^{j\,l} \bigr\rangle
\nonumber \\ &&
+\frac{4\,i}{M}\,\sum_a
\biggl[\frac{e_a}{m_a}\biggl(\frac{g_a}{3}+1\biggr)+\frac{Z\,e}{M}\biggr]
\bigl\langle Q^i\,{\rm Ln}\,x_a^i\,s_a^k - {\rm h.c.}\bigr\rangle
\nonumber \\ &&
+\frac{2\,i}{M}\,
\sum_a\,\biggl[\frac{e_a}{m_a}+\frac{3\,Z\,e}{4\,M}\biggr]\,
\bigl\langle Q^i\,{\rm Ln}\,(l_a^k\,x_a^i+x_a^i\,l_a^k)
-{\rm h.c.} \bigr\rangle
\nonumber \\ &&
-\frac{16}{M^2}\,\epsilon^{ijk}\langle D^i\,{\rm Ln}\,Q^j\rangle
-\frac{14}{M^2}\,\langle \vec D \times \vec Q \rangle^{\,k}
+\biggl(\frac{Z\,e}{M}\biggr)^2\,S^k\,\biggl[\frac{3}{2}\,g^2 + 26\,g-10\biggr]
\nonumber \\ &&
-\sum_a \biggl(\frac{e_a}{m_a}\biggr)^2\,\langle s_a^k\rangle
\,\biggl[\frac{3}{2}\,g_a^2 + 26\,g_a-10\biggr]\biggr\}
+\sum_a \frac{e^2}{6\,m\,m_a}\,\psi^2(0)\,\vec\sigma\langle\vec s_a\rangle\,\delta g_a,\label{68}
\end{eqnarray}
where ${\rm Ln} = \ln[2\,(H_{\rm IN}-E)/m]$, 
$m$ is the electron mass, $\delta g$ is defined in Eq. (\ref{59}), 
$D^i$ in Eq. (\ref{23}), $\mu^i$ in Eq. (\ref{26}), 
$Q^i$ in Eq. (\ref{29}), $Q^{ij}$ in Eq. (\ref{33}).
Moreover, our definition of the nuclear $g$-factor Eq. (\ref{02}) 
differs from the standard one by the use of the actual nuclear charge and mass,  as
opposed to the unit charge and the proton mass. 
The above expression is quite complicated, and its use may be limited by the
lack of an accurate nuclear wave function. However, 
it takes a particularly simple form for the deuteron,
\begin{eqnarray}
\delta E_{\rm pol} &=& 
\frac{e^2}{64\,\pi^2}\,\psi^2(0)\,\sigma^k\,\biggl\{
-12\,i\,\epsilon^{ijk}\,
\langle e\,R^i/2\,\ln[2\,(H-E)/m]\,e\,R^j/2\rangle\nonumber \\ &&
-S^k\,\biggl[\frac{e\,(g_p-g_n)}{2\,m_p}\biggr]^2\,\langle\,\ln[2\,(H_S-E)/m]\,\rangle
\nonumber \\ &&
+\frac{4\,S^k}{m_p}\,\biggl(\frac{g_p-g_n}{3}+1\biggr)\,
\langle e\,R^i/2\,(H_T-E)\,\ln[2\,(H_T-E)/m]\,e\,R^i/2\rangle
\nonumber \\ &&
+S^k\,\biggl(\frac{e}{m_p}\biggr)^2\,\biggl[
\frac{1}{4}\,\biggl(\frac{3}{2}\,g_d^2 + 26\,g_d-10\biggr)
-\frac{1}{2}\,\biggl(\frac{3}{2}\,(g_p^2+g_n^2) + 26\,g_p-10\biggr)\biggr]
\biggr\}\nonumber \\ &&
+\frac{e^2}{12\,m\,m_p}\,\psi^2(0)\,\vec\sigma\,\vec S\,(\delta g_p + \delta g_n),
\label{69}
\end{eqnarray}
where $\vec R = \vec r_p-\vec r_n$ and $H_T$ and $H_S$ are nonrelativistic
proton-neutron Hamiltonians in the  triplet and singlet states
respectively. Further on, we used an approximate deuteron mass $m_d = 2\,m_p$
and neglect $D$-wave mixing of the ground state.

The nuclear structure correction to hfs in deuterium, including some polarizability
effects, has been obtained
by Khriplovich {\em et al.} in Refs. \cite{khr1, khr2}. Their result of $153$ ppm  seems to explain
well the difference between the experimental value and theoretical predictions with
assuming a point nucleus, which is  $138$ ppm, see Eq. (\ref{04}). 
We point out,  however, that their calculations were performed
only in logarithmic accuracy, and in our result the
$\ln\Lambda$, the logarithm of an arbitrary chosen cut-off cancels out. 
Friar and Payne have recently calculated in Ref. \cite{friar} $\delta E_{\rm Low}$ and $\delta
E_{\rm Zemach}$. Their result for deuterium of $141$ ppm accounts for most of the 
difference in Eq. (\ref{04}), but they have not included $\delta E_{\rm pol}$,
which they claim to be negligibly small, in disagreement with our findings.
It would be now interesting to calculate the nuclear polarizability correction to  hfs 
according to Eq. (\ref{69}) and to 
verify it against precise experimental value for hfs in deuterium \cite{deut} in Eq. (\ref{03}),
as well as in other light atoms. As it is highly difficult to go beyond
$\delta E_{\rm pol}$, the value of this correction sets the ultimate limit of the accuracy 
theoretical predictions for light atomic systems, which we expect to be a few ppm,
as they are for hydrogen hfs \cite{carl}.

The nuclear polarizability correction $\delta E_{\rm pol}$ is of order
$O(Z\,\alpha\,m/m_p)$, thus, in general, it is  smaller than  
$\delta E_{\rm Low}$ or $\delta E_{\rm Zemach}$. However, in some cases
we expect this correction to be more significant. When a closely
lying state of the opposite parity is present,
such as in $^{11}$Be nucleus, with a large $B(E1) \approx 0.1\;e^2\,{\rm fm}^2$
line-strength \cite{11Be}, 
$\delta_1 E_{\rm hfs}$ in Eq. (\ref{25}) can be as large as 
$B(E1)\,m\,m_p\,E_F\approx 10^{-4}\,E_F$, which should be possible to 
verify experimentally. In fact, the measurement of the hyperfine splitting
in $^{11}$Be$^{+}$ has already been proposed in Ref. \cite{nak} for the determination
of the  neutron halo. We do not think that 
the contribution to hfs from the neutron halo can easily be identified,  
but this measurement can verify the significance of the nuclear vector
polarizability in the hyperfine splitting of atomic systems.

\end{document}